# Framework for Behavioral Disorder Detection Using Machine Learning and Application of Virtual Cognitive Behavioral Therapy in COVID-19 Pandemic

Tasnim Niger [1] [*], Hasanur Rayhan [2], Rashidul Islam [3], Kazi A. A. Noor [2], Kamrul Hasan [1]

[1] *Computer Science & Engineering, Islamic University of Technology, Gazipur, Bangladesh;*
[2] *Computer Science and Engineering, Shanto-Mariam University of Creative Technology, Uttara, Dhaka, Bangladesh;*
[3] *Computer Science and Information Technology, Shanto-Mariam University of Creative Technology, Uttara, Dhaka, Bangladesh*

In this modern world, people are becoming more self-centered and unsocial. On the other hand, people are stressed, becoming more anxious during COVID-19 pandemic situation and exhibits symptoms of behavioral disorder. To measure the symptoms of behavioral disorder, usually psychiatrist use long hour sessions and inputs from specific questionnaire. This process is time consuming and sometime is ineffective to detect the right behavioral disorder. Also, reserved people sometime hesitate to follow this process. We have created a digital framework which can detect behavioral disorder and prescribe virtual Cognitive Behavioral Therapy (vCBT) for recovery. By using this framework people can input required data that are highly responsible for the three behavioral disorders namely depression, anxiety and internet addiction. We have applied machine learning technique to detect specific behavioral disorder from samples. This system guides the user with basic understanding and treatment through vCBT from anywhere any time which would potentially be the steppingstone for the user to be conscious and pursue right treatment.

Behavior can be defined the way which an individual behaves or acts.[1] It is the way an individual acts towards people and society in different context. Behavioral disorders relate to mental health problems that lead to disruptive behavior, emotional and social problems.[2]

Since the rise of COVID-19 pandemic and subsequent lockdowns, human being is becoming more isolated and self-centered.[3-7] Particularly the students of the educational institutes who are bearing the brunt.[8] The lockdown is impacting social life & connection and thus leading to become more reliant on technology.[6] Social distance in COVID-19 time creating disorders like depression, anxiety, internet addiction etc.[9] The paper intents to work on the mental disorders of the university students and to find a solution to detect the likelihood of behavioral disorders.

To detect the right disorders, mental health professionals have long & multiple sessions with the individuals. Through structured questionnaire professional tried to detect the right disorder. Sometime the outcome becomes time consuming and not effective. So, we built a digital framework with right test attributes for collecting data to detect the right behavioral disorder using machine learning technique.

Machine-learning algorithm is a program that adjusts its own parameters, feedback on the experience that it gains beforehand from the provided dataset.[10] The user can feed a machine learning algorithm with massive quantity of data to analyze and render data-driven decisions and recommendations based on the input data. System can learn through data or observation usually in the form of instruction, example or direct experience in which the application looks for patterns that match. According to the found pattern, the computer tries to give better result over time with new dataset.

To perform detection using machine learning the more sample we give and the more experience that it gets from the users, the better the prediction will be. That will allow our application users to get an idea about their mental state on whether they are in any crisis condition and receive initial support.

By leveraging patient data, machine learning-aided diagnosis is set to revolutionise cognitive healthcare by providing precise and personalised diagnoses. Clinically useful classifiers can distinguish diagnoses from multi-dimensional datasets. For the research purpose we have collected data set and fed to the machine learning model, we identified one thousand subjects and surveyed them as a representative sample for the following problems: Depression, Anxiety and Internet Addiction.

Depression and anxiety are two very common behavioural disorders. Although there is certainly a bit of overlapping between the two disorders. However, the detection of the patient behavioural disorder is very important to ensure adequate treatment. While anxiety is the concern of what might be happening where depression in cases is the outcome of anxiety. An individual with anxiety disorder feels drained after a period of intense anxiety whereas depression patient feels constant fatigue.[11] During COVID-19, internet is the most important part of our lifestyle. But this interaction with information technology is leading to issues. Internet addictions create strong dependence on internet leading to depression, anxiety, stress etc.[9,12] Excessive usage of internet can be uncontrollable and heavily impact the student's behavioural pattern.[12]

Upon completion of the first phase DETECTION and after user confirmation user will be redirected to a sample website according to the problem detected by our application. That website would give the users the basics of virtual CBT (vCBT). There will be links on the website which would redirect the user to some audio-visual first aid therapy that might give the initial support before getting professional help.

**Behavioral Disorder**

Behavioral disorder can be patterns of multiple disruptive behaviors which effects human life.

According to the Individuals with Disabilities Education Act[13] Behavioral Disorder or Emotional Disturbance is a condition displaying some characteristics over an extended period that unpleasantly affects a person's emotion and performance. Persons may have inability to learn, build or maintain social, interpersonal relationships with people. They can behave inappropriately under normal circumstances. Always possess unhappiness or depression mood.

*Specific Behavioral Disorder chosen for this research*

- Depression - Depression is a very common and serious behavioral disorder. Depression is highly prevailed amongst the population hence considered as very common.[14] Depression takes away all the pleasure of personal to social life ranging from eating, sleeping, resting, playing, studying, mental capacity to analyze things, entertaining etc.[15]

- Anxiety - Anxiety itself is not a natural behavioral disorder as every living being feel anxiety which is standard and often can be seemed like healthy emotion.[14] However, when the feeling gets more frequent and can unearth the person with highly disproportionate means that disrupt regular day-to-day life, then it can be considered as a mental disorder that needs to be looked after. Therefore, if a person often feels unbalanced levels of anxiety, it might become a health disorder.[16] Usually the diagnoses are as follows: excessive nervousness, fear, apprehension and feeling of extreme worry. These feelings may alter how emotion is processed and thus, effecting external behavior which may cause physical symptoms. Minor anxiety might be elusive and disturbing, but in case of severe disorder, it may extremely affect a person's everyday living.

- Internet Addiction - It has become a common behavioral disorder among adults in this digital era and usually have a negative effect on one's intellect and health.[17-19] The severity of its impact depends on people's behavioral addiction level. People occupy in a behavioral pattern frequently despite it causes them long-term harm. At the end it becomes complex and difficult to overcome this addiction.

The likelihood of these different pattern of behavioral disorder can be detected and therapy can be provided through virtual cognitive behavioral therapy (vCBT).

**Cognitive Behavioral Therapy (CBT)**

Cognitive behavioral therapy (CBT) is a helping process to overcome person's emotional and intellectual experience: how they are feeling and what they think about the problem they have sought for help.

CBT is a hands-on, practical method of psychotherapy that is implemented for a smaller period to reach to the goal faster and approach to self-healing/thinking approach so that the patient alters his mental situation by his own core beliefs. The primary goal is to change how a person thinks or behaves that causes the difficulty or mental block. It is used to treat a wide range of issues in a person's life. CBT aims to change one's attitude and behavior towards life by focusing on the feelings, norms, beliefs, and attitudes.[20]

*Steps in CBT*

Cognitive behavioral therapy (CBT) is customized to cater patient's definite condition. CBT process, techniques and required steps need to be understood for successful implementation.[21] The steps in CBT are given below:

(1) Identify disturbing circumstances or situations of life and define focus areas.
(2) Awareness of feelings, sentiments, and beliefs. Usually, patients are prescribed to record of every thought.

(3) Identify negative or erroneous thought process. Advised to be concerned about every physical and behavioural responses to every circumstances.
(4) Restructure negative or erroneous thinking process which takes time and patience.
(5) CBT is usually treated as a short-rang therapy which last about 10 to 20 sessions.

*Effectiveness of CBT*

CBT facilitates positive changes in person's life. It is a psychotherapeutic approach of helping someone to come to terms with and workout solution to their problems. However, they vary based on the approach used and underpinning model according to the category of disorder.

Depression – One of the Australian cohort study identified that CBT and pharmacological treatments had reasonable impact on Depression.[22] When suffering from severe depression, one requires long-term treatment strategies that involve either medication or psychological therapy.[23]

Anxiety - A meta-analysis by Kristin Mitte evaluated the effectiveness of cognitive behavioral therapy and pharmacological therapy in treating comprehensive anxiety disorder. There were 869 patients involved in that analysis. The analysis includes thirteen studies which compares CBT to regulate the groups and six comparing CBT with pharmacotherapy. She concluded that, the present meta-analysis results specify CBT is a highly effective treatment that reduces the key signs of anxiety as well as the related depressive symptoms. As a result, life quality improves subsequently.[24]

Internet Addiction - Addictive behavior has a wide range of variety starting from drug abuse to the minimum of mobile phone addiction and each addiction is detrimental in their own way and cannot be undermined. Professor of Psychology, Charles Scherbaum casually divided 73 opioid-dependent patients into two groups. He gave methadone to the first group with 20 weeks of group CBT.[25] On the other hand, second group received methadone and treatment as usual i.e counselling. Neither group had a different rate of negative urine tests after weeks of treatment. However, the CBT group had a lower rate of positive urine drug tests after months of follow-up.[26]

## Materials and methods

**Proposed vCBT Framework**

Our purpose was to figure out a way to detect if a subject is suffering from either one of the followings:

-Depression
-Anxiety
-Internet Addiction

There are several classic machine learning algorithms out there that can help us with this quest and some of these algorithms can be directly implemented in our study to get the result that we are looking for.

**Assessment Methodology**

The outcome of Supervised Learning can be formulated as a function of the features (given) so that model classify the exact outcome when features are given in the future.[27] Commonly used examples of Supervised Learning algorithms include SVM and KNN. SVM is most widely used in psychiatry.[28] SVM has generalization capabilities which prevent it from over-fitting. In our case, our dataset needs normalization and transformation, SVM was the perfect choice. We found multiple applications of SVMs for two class and multi class classifications which was perfect as we needed three-dimensional classification. A small change to the data does not greatly impact the hyperplane and hence the SVM. So, the SVM model is steady.[29]

In KNN, no computations are required before actual classification because KNN is instance-based learning.[30] Data given in the model considered as 'learnt' model and computation performed at the time of actual prediction. Label data can be detected from similar nearest neighbour data points.[31]

**Framework design**

At first, we opted for surveying the students. Respondents entered data using our designed website form. Data needed to be normalized and resolved as there were missing values, unwanted value or values that made no sense. Next, we applied SVM and KNN algorithm, calculated accuracy of each algorithm and selected the right one. We created a static website that contained the interface for the users. Later we converted the static site to a desktop app. According to the generated model user received the result. The user was redirected to the site for self-help after agreeing with result.

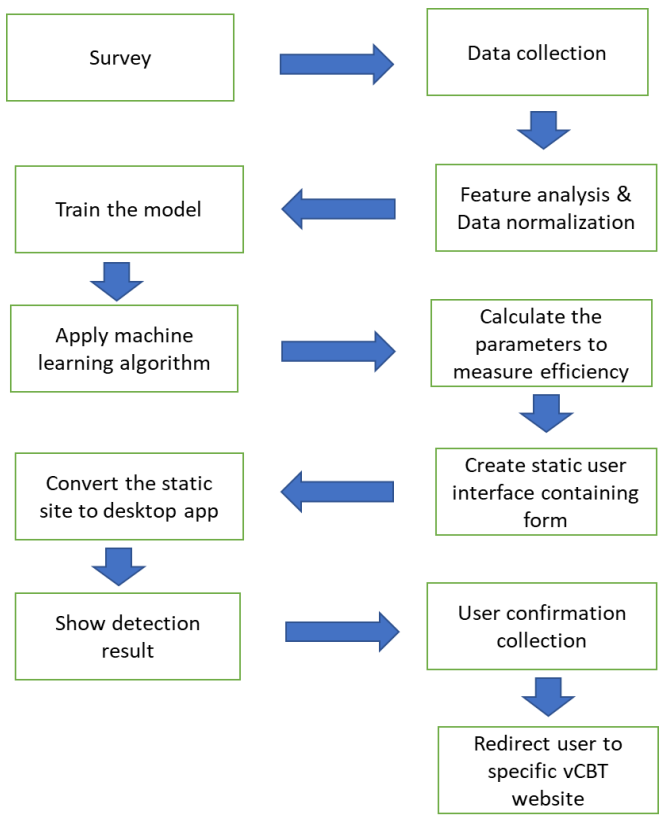

Figure 1. — *Design of framework*

**Data Collection**

Machine learning focuses on the implementation of computer programs to access data and automatically learn by using it.

Our first step was to collect enough data with adequate columns to train our model sufficiently as ours is a supervised learning algorithm. It was not always easy to obtain dataset in pandemic situation. We prepared a questionnaire and designed in a specific way to collect enough information for a complete study. We studied only on students from our university of different departments and tried to figure out their mental state with the set questionnaire. Most of our questions were closed question. Closed question gives specific information – it must be answered with either a single word or a short phrase. Questions were prepared using easy text. So that, user can understand easily and answer without hesitation.

*Data Collection Method*

Our data collection process had structured dialogue with our target group i.e. university students. There were four ways of data collection – in-person interviews, online interview, mail, and phone call. From the above four, we chose mail, phone call and online questionnaire considering the COVID-19 situation. Our target was to get primary data that is reliable and can be easily used in our machine learning algorithm for better analytics. The term "primary data" refers to data anyone gather by themselves, rather than the data collected after another party originally recorded it. So, primary data is the most reliable that can be collected directly from the source. Protecting confidentiality is imperative to maintain the trust of data providers. Protecting participants is critical in the form of preventing psychological damages such as embarrassment or distress. Maintaining privacy and confidentiality helps to avoid social harms such as job loss or financial damage.

We used to create an interface that would contain of a form which user may fill. Once the users submit the form the page will redirect to a page that will tell him/her about the probable mental condition. The user will then provide his/her consent. Once agreed he/she will be redirected to page of virtual CBT. The interface that we designed was a web interface, but we made a desktop app for the successful conversion.

A screenshot of the interface is given here:

Figure 2. — *Web Interface to collect data*

This is a web-based form with input tags. Each element is identified with ID's which are later used to fetch the input from the interface.

*Data Collection Domain*

Our primary target was to find out behavioral anomalies of university students. We decided to keep our domain of data collection only with our university students. This enabled us to get information from different socioeconomic class of students from different semesters. We surveyed 1000 students at the university.

*Questionnaire Analysis*

We set the questions that would be direct fit to our framework. Some of our questions were too direct just to rattle the respondents which may make them exposed and further would give us an indication of his/her mental state.

*Sample Dataset*

Our target was to equipped dataset with two kinds of data.[32]

- Information relating to students' background aptitudes, intelligence, achievements, interest, plans etc.
- Data about the areas in which students may seek assistance. Those areas may be personal and educational.

Table I. —*Analysis of some sample features*

| Sample | Detail |
|---|---|
| Age | Most of the subjects are within the age of 18-25. However, the average is 23 years |
| Number of Subjects | We have considered 1000 students for the analysis |
| Sex | Most of the students are male. The female students are around 22% |
| Departments | We have considered students from multiple departments like CSE, CSIT, GDM, English, Architecture etc. |
| Employment status | Employment status was also considered. Out of the 1000 students 489 were employed. |
| Social Status | We have taken representation from all the social segments from poor to rich with almost equal split |
| Chronic Disease | 162 of the 1000 students has chronic diseases |

**Feature Engineering**

The numbers of features are 18. Following are the features (with options) used to collect data:

- ID: Auto generated
- Age: Integer
- Sex: 1 for Male, 0 for Female
- Literacy: 1 for Literate (Minimum Higher Secondary School or equivalent), 0 for Illiterate
- Marital Status: 0 for Married person, 1 for Unmarried person, 3 for divorced
- Children: No - if no children, yes for children
- Employed: 1 for employed, 0 for unemployed
- Socio Economic Status: (1-5, 1 being poor and 5 being self-sufficient)

- Drug Addiction: 1 for yes, 0 for no
- Chronic Disease: 1 for yes, 0 for no
- Medication: 1 for yes, 0 for no
- Education: (1-5, 1 being primary diploma holder, 2 HSC or 'A level', 3 B.Sc., 4 Masters and 5 PhD )
- Financial Status: (1-10, 0 being poor, 1- no income, 2- low income, 3- in loss situation , 4- no loss no profit, 5- moderate income, 6- sufficient income, 7- High income, 8- established , 9- rich, 10 highly rich)
- Income: amount in Bangladeshi Taka
- Sleeping Hour: (0-24) hr.
- Result Satisfaction: 1 for yes, 0 for no
- Feelings of overwhelm: 1 for yes, 0 for no
- Extracurricular activities: 1 for yes, 0 for no
- Hangout with friends in hour:  No or (1-10, in hours)

Labels were encoded using some numeric values. Based on the inputs - data features may determine the specific disorder. If someone give inputs that entails pain/discomfort, chronic disease or drug seeking behavior or may showcase family problem – our model may assess the class indication as "*depression*".

Similarly, if someone has trouble falling in sleep, feel fatigue, have impaired concentration and always far of social performance/extra curriculum activities – the output indication may be termed as "**anxiety**".

In case where we see less concentration, unemployment, sleep trouble, divorced due to lack of social interaction or relationship - such assessment may indicate "**Internet addiction**".

So, the data features is framed with specific inputs which disseminate wide varies of data from where the detection of disorder is possible.

*Resolve Missing Values*

We observed very few missing values during our feature analysis phase. Those missing values were replaced by using mean or mode. We used mode value for categorical data and mean value for numerical data to replace missing values.

*Specify Target Value*

We had 3 targets. Those were Depression, Internet Addiction and Anxiety.  We had used the following values of our 3 targets:
- Depression          : 1
- Internet Addiction  : 2
- Anxiety             : 3

*Data Transformation*

For better accuracy and to fit KNN and SVM algorithms with our dataset, we made further data normalization on our dataset. We collect the data in some cases from range 1-10 or 1-5 but that needed to change for better precision and accuracy. We applied min-max normalization technique and make the highest range 1 and the lowest 0. So, in this case, where the previous data was 5, now it has become 0.5 and so on. This was a necessary to get the desired result. To convert categorical data to numerical data we have applied label encoder.  Like we transformed 'female' to 0 and 'male' to 1. Similar for literacy and marital status features and target column. During normalizations process, through trial-and-error method, we changed, normalized column in the process of SVM.

**Splitting Dataset and Apply Machine Learning Approach**

To prevent overfitting, we applied 10-fold cross-validation on 1000 student's data. We split total amount of data into 80% (training dataset) and 20% (testing data set). There are 100 elements in each fold. We then ran the train data through the classifier. Upon completion of training, we used two machine learning methods SVM and KNN to create the models. In SVM, a hyperplane is selected to best separate the points in the input variable space by their classes, either class 1, class 2 or class 3. In Support Vector Classier (SVC) we used linear kernel, set gamma to 0 and C parameter to 1 to classify the training point correctly. In KNN, set neighbours parameter to 3. The models would further allow us to test user data and according to the model state, the user's result will be shown. Each time, we run the test file, models are generated according to our data structure. These models are responsible for the test file to run smoothly and accordingly. Finally, the test file generates the result using the model.

**Detection**

Our system is totally a person-centred therapy. Upon detection of Depression, Internet Addiction or Anxiety, system generates an output that is formatted as either 1, 2 or 3.

Detection of the right disorder from our framework is not 100% accurate. However, it will provide a close indication of the behavioural disorder. The user will then agree on the outcome in a separate form. Once agreed depending on the result, our application redirects the user any one of the three sample websites. From those websites' users can be guided according to prescribed therapy.

**Virtual Therapy through External Website**

Virtual psychotherapy system plays non-judgemental attitude to detect the correct disorder. Whereas it is stressful for a human therapist to maintain genuineness, warmth, accurate empathy, respect and non-judgemental attitude.[33]

The first step to solve any sort of disorder related problem the individual needs to step out and accept that he/she has a problem, and this can be diagnosed by our vCBT. The mind set needs to be clear and honest to rely on vCBT.
This vCBT encourages the user to go back to their earlier experiences and realize how these experiences affect their current 'problem'. Self-awareness arises among the users. Self-awareness is the ability to look inside and recognize their own psychological needs and desires, such as a need for concentration on their education, background, religious or political beliefs or desire to be professionally competent. In second step prescribed contents of the website guided the users how to control their emotions especially of anger, impatience, and frustrations. The users of our system get the wide range of therapy including relaxation or anxiety-control training, self-management procedures and virtual guidance like reading story book, listening to music, physical exercise etc. Users find the best interest using this method and system received the trust of the users.

Following are the second steps vCBT for depression, Internet addiction and anxiety.

*Second step vCBT for depression*

- Music therapies to reduce stress.
- Post of job circular to get a new job.
- Reference of support groups in local community for joining.
- Provide help to involve in physical exercise through videos/ trainer's advice. This will boost serotine, endorphins which triggers the growth of new brain chemicals and connections.
- Guide for healthy food and lifestyle by the expert.
- Professional advice on antidepressants medication.

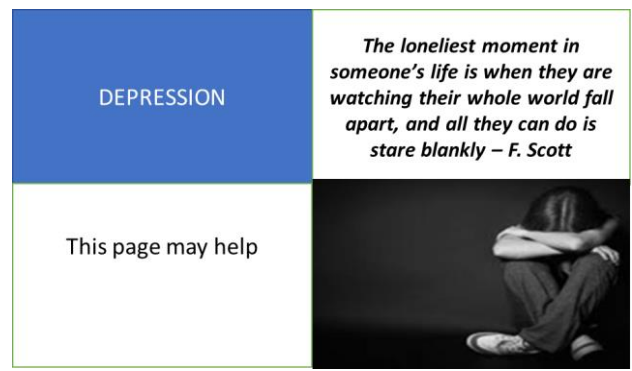

Figure 3. —*Depression Website Layout*

*Second vCBT for internet addiction*

- Encourage to enjoy more time with family and friends.
- Set rules to restrict/ curve the usage of internet after a certain time each day.
- Set up rules on device to limit online session to 30 minutes.
- Information and guidance on travel & fun activities to enjoy with friends, relatives and loved ones.
- Guidance on digital priority to focus on study and core deliverables.

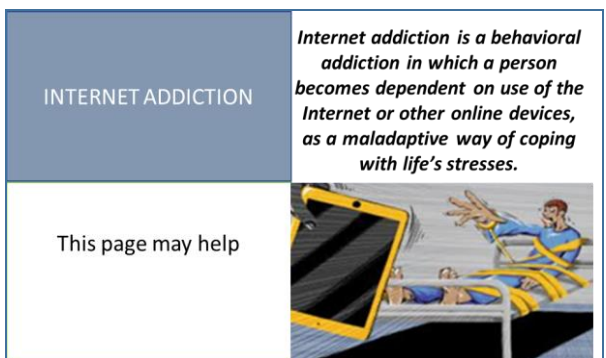

Figure 4. —*Internet Addiction Website Layout*

*Second step vCBT for Anxiety*

- Refer good and impactful book and sourcing.
- Short E-learning course reference.
- Motivational therapy guidance to transform negative thought to positive thoughts.
- Relaxation therapy such as mindfulness meditation or yoga and progressive muscle relaxation.

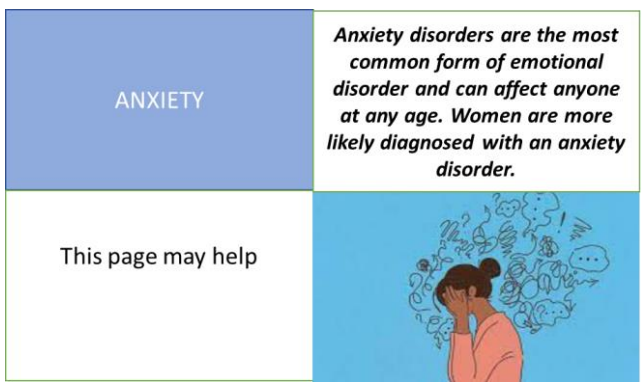

Figure 5.— *Anxiety Website Layout*

# Result

To measure the classification performance, we have used the term Precision, Accuracy, Recall and F1-score.

**Precision** - Precision (positive predictive value) is the ratio of relevant positive observations to the total retrieved positive observations.[34]

**Recall** (Known as Sensitivity) - Recall is the ratio of relevant positive observations among all observations in actual class - yes.

**Accuracy** - In any measurement of performance, accuracy is a ratio between relevant observations and the total observations. We can say our model is best if we've high accuracy. Accuracy may be a great measure but only when values of false positive and false negatives are almost same of symmetric datasets.

**F1 score** - Weighted average of precision and recall is called F1-score. Therefore, F1-score consider both false positives and false negatives. Naturally it is not as easy to achieve as accuracy, however if there is an uneven class distribution F1 is generally more useful than accuracy.[35]

In our case, we see that in the support column, more than 60 percent of our test subjects noted that they are depressed, about 25-30% said they are addicted to internet and the rest are anxious. Accuracy only works better when the data is even, but that is not our case. So, our standard of efficiency should be the F1-score.

We achieved accuracy, precision and F1-score up to 80 percent using SVM (Table 1).

Table II. —*Accuracy result using SVM*

|  | Precision | Recall | F1-score | Support |
|---|---|---|---|---|
| 1 | 0.86 | 0.90 | 0.88 | 60 |
| 2 | 0.88 | 0.70 | 0.78 | 30 |
| 3 | 0.38 | 0.50 | 0.43 | 10 |
|  |  |  |  |  |
| Accuracy |  |  | 0.80 | 100 |
| Macro avg | 0.71 | 0.70 | 0.70 | 100 |
| Weighted avg | 0.82 | 0.80 | 0.80 | 100 |

Table III. —*Accuracy result using KNN*

|  | Precision | Recall | F1-score | Support |
|---|---|---|---|---|
| 1 | 0.93 | 0.90 | 0.92 | 62 |
| 2 | 0.74 | 0.82 | 0.78 | 28 |
| 3 | 0.44 | 0.40 | 0.42 | 10 |
|  |  |  |  |  |
| Accuracy |  |  | 0.83 | 100 |
| Macro avg | 0.71 | 0.71 | 0.71 | 100 |
| Weighted avg | 0.83 | 0.83 | 0.83 | 100 |

The weighted average of our f1-score using KNN (Table 2) is 83 which is an acceptable result and better than F1 score of SVM. Hence, we decided to use KNN for our framework.

## Discussion

The purpose of the paper was to detect the physiological disorder of university students during this Covid-19 time. Two different machine learning methods were used to detect the disorders. First, we chose SVM as our machine learning technique since it looked like the more feasible choice. Using SVM, we had a precision and accuracy of about 60 percent at the start and we need to make a complete makeover of our dataset and normalized it to get the accuracy and precision up to 80 percent.

As we were thinking about making a desktop app that takes input from an interface, we tried different approaches and we had to renormalize our data and made some major changes. We changed our model from SVM to KNN because the new normalized data set was best suited for KNN. SVM can easily separate dataset using decision planes, i.e. the basic SVM separate classes using linear hyperplanes but in the cases of different kernel, it will change the shape of decision plane. K-Nearest Neighbour (KNN) works better than SVM. KNN can generate highly complex decision as the main driving issue is the raw training data. SVM uses a limited parametric estimate of the decision limit, which is fine if anyone prefer performance against processing speed ratio.

In our case, our classes are quite easily separable and that is a precondition for KNN to provide good results; if KNN failed, then it would have pointed to the fact that the metric vector we have chosen does not produce separable classes and that is quite opposite to our dataset. So, KNN gave us the better accuracy and precision with the current dataset that we have.

However, in future we will apply different types of decision tree methods to get better accuracy.

## Limitation

One of the limitations of the exercise was limited number of features. Few specific special features were selected during feature engineering step to build the model and to fasten the process. Also, another limitation was sample data. The overall sample data was in lower base due to pandemic situation

## Conclusion

Proper use of machine learning and modification of some of the algorithms may produce satisfactory result in determining mental disorder given a complete and information rich dataset. The Implementation part above shows that our model yields satisfactory result, and the precision is also reasonable. The algorithm we used on our dataset can be tweaked and twitched further for better result in the future.

About the dataset, our data is one dimensional as all our test subjects were students. So, there are room for improvement. The more dimension that we can add in our dataset, the tougher the calculation would have been, but at the same time, the result could have been more precise.

As we gave precedence in detecting target value out of the given input, making a desktop interface, and connecting external websites for the first step as self-help was secondary in our eyes. This can further be improved given enough time, adequate study on virtual CBT, manpower and willpower. We would also need specialist on CBT who can guide us through our interface making process to ensure accurate helps that can be done for the people. But for the first step to wellness, we think we have given our users enough chance to further study on their probable condition which may lead to their betterment.

Finally, the final step of this project would be to publish it on cloud platform being a one stop, cross platform solution for people with behavioral disorder.